# A Color Image Digital Watermarking Scheme Based on SOFM

J. Anitha[1], S. Immanuel Alex Pandian[2]

[1]Asst.Prof., Dept. of Computer Science and Engineering, Karunya University,
Coimbatore, Tamilnadu, India.

[2]Asst.Prof., Dept. of Electronics and Communication Engineering, Karunya University,
Coimbatore, Tamilnadu, India.

**Abstract**
Digital watermarking technique has been presented and widely researched to solve some important issues in the digital world, such as copyright protection, copy protection and content authentication. Several robust watermarking schemes based on vector quantization (VQ) have been presented. In this paper, we present a new digital image watermarking method based on SOFM vector quantizer for color images. This method utilizes the codebook partition technique in which the watermark bit is embedded into the selected VQ encoded block. The main feature of this scheme is that the watermark exists both in VQ compressed image and in the reconstructed image. The watermark extraction can be performed without the original image. The watermark is hidden inside the compressed image, so much transmission time and storage space can be saved when the compressed data are transmitted over the Internet. Simulation results demonstrate that the proposed method has robustness against various image processing operations without sacrificing compression performance and the computational speed.

## 1. Introduction

With the rapid development of multimedia and the fast growth of the Internet, the need for copyright protection, ownership verification, and other issues for digital data are getting more and more attention nowadays. Among the solutions for these issues, digital watermarking techniques are the most popular ones lately [1]–[5]. Digital watermarking techniques provide effective image authentication to protect intellectual property rights. The digital watermark represents the copyright information of the protected image.

Digital watermarking is the process of embedding a secret information (i.e., a watermark) into a digital data (namely audio, video or digital image), which enables one to establish ownership or identify a buyer. Digital watermark can be a logo, label, or a random sequence. In general, there are two types of digital watermarks, visible and invisible watermarks. The embedded watermarks can later be extracted or detected from the watermarked media for authentication or identification. Invisible watermarks can be broadly classified into two types, robust and fragile watermarks. Robust watermarks are generally used for copyright protection and ownership verification because they are robust to nearly all kinds of image processing operations. In comparison, fragile watermarks are mainly applied to content authentication and integrity attestation because they are completely fragile to any modifications.

In general, the digital watermark technique must satisfy the following properties. First, the embedded watermark does not distort visually the image & should be perceptually invisible. The second property is that the watermark is difficult for an attacker to remove. It must also be robust to common attacks.

Until now, many researchers have come up with a variety of watermarking techniques, and the techniques proposed so far have been highly effective in the area of copyright protection. However, most current watermarking techniques embed the watermark in a normal, original image instead of a compressed image [7]. They offer considerable security for normal, uncompressed images, and yet they fail to take into consideration the inevitable fact that, due to limited frequency bandwidth, almost all of the images travelling over the Internet are in compressed forms in order to save transmission time and storage space. Therefore, to allow for application to the Internet, a new watermarking approach has recently been used to combine with image compression techniques to embed a digital watermark into the compressed image [8], [9]. It is worthwhile to note that image compression and image watermarking have some common features.

In image compression [16], an image is encoded using its most significant characteristics. After compression, the greatest concern is that the distortion of the decompressed image must be controlled. On the other hand, in image watermarking, the embedded watermark must not degrade the image quality of the watermarked image. With this similar goal, the image compression approach can aid in the implementation of watermark embedding and detection. Recently, Lu et al. (2000) [6] presented a VQ-based digital image watermarking method. This method utilizes the codebook expansion technique to produce a partition of the codebook. According to the different sets in the partition of the codebook, the watermark can be embedded into an image. This method is simple and effective. The authors claim that if the codebook is kept secret, their method can provide the secure feature. Makur and Selvi (2001) [10] proposed another watermarking scheme based on variable dimension VQ (VDVQ) for hiding a watermark into images. The main property of VDVQ is the use of a codebook that contains codewords of varying dimensions. To embed the binary watermark, this method uses a codebook that contains codewords of two different dimensions. When an image block is encoded, the dimension of the selected codeword is set according to the corresponding embedded watermark bit. However, the embedded watermark is fragile. The decoding and re-encoding of the compressed image may easily remove the embedded watermark. Hsien-Chu Wu and Chin-Chen Chang proposed another watermarking scheme based on codebook editing in LBG algorithm [11], [12].

In this paper, we propose a new method of digital watermarking that embeds the watermark in an already compressed color image through VQ technique [13]. Here the SOFM based codebook generation algorithm is used. The watermark is embedded in the selected index of the codebook. In color image, R channel is used for watermark embedding. The watermark exists both in compressed and reconstructed image. The experimental results show that this method enables images to withstand a number of destructive image processing procedures including cropping, blurring, sharpening and JPEG compression [17].

## 2. SOFM based VQ

Codebook plays an important role on every compression technique. This codebook is used for both compression and a decompression stage, which is generated by means of SOFM based VQ [14], [15]. A vector quantizer Q of dimension k and size S can be defined as a mapping from data vectors in k-dimensional Euclidean space, $R^k$ into a finite subset C of $R^k$. Thus

$$Q : R^k \rightarrow C$$

Where C={$y_1, y_2, \ldots, y_s$} is the set of S reconstruction vectors, called a codebook of size S, and each $y_i \in C$ is called a code vector or codeword. For each $y_i$, $i \in I,\{1,2,\ldots,S\}$ is called the index of the code vector and I is the index set. Encoding a data vector $x \in R^k$ involves finding the index j of the code vector $y_j \in C$ such that $\| x - y_j \| \leq \| x - y_i \| \; \forall \; i \neq j$ and $i,j \in I$. The decoder uses the index j to look up the codebook and generates the reconstruction vector $y_j$ corresponding to x. the distortion measure $d(x,y_j) = \| x - y_j \|$ represents the penalty of reproducing x with $y_j$. If a VQ minimizes the average distortion, it is called the optimal VQ of size S.

A self organizing feature map is a neural network clustering technique, having several desirable features in the field of VQ. The learning scheme of the self organizing feature map is an application of the least mean square algorithm where the weights of the neurons are modified on the fly for each input vector. Thus the codebook is updated using an instantaneous estimate of the gradient known as stochastic gradient, which does not ensure monotonic decrease of the average distortion. Also due to incorporation of neighborhood update (opposed to the winner only update in pure competitive learning) in the training stage, SOFM networks exhibit the interesting properties of topology preservation and density matching.

The SOFM algorithm generates the weight vectors. The resultant weight vectors are called the SOFM codebook. The codebook obtained from the SOFM algorithm is used for codebook partition to perform watermark embedding.

## 3. Codebook Partition

Our watermarking technique is mainly depends on the codebook partition to complete watermark embedding process. Let the initial codebook trained by SOFM be CB such that each codeword CW in CB contains a*a elements. After the codebook partition process, in each division, there will be two codewords, numbered 0 and 1, respectively, so that codeword 0 is used to embed watermark bit 0. The codeword 1 corresponds to watermark bit 1. The two codewords most similar to each other are classified into the same division with their index values recorded in the codebook. The degree of similarity between one codeword and the others is determined by calculating the mean square error (MSE). The MSE of two codewords CWx and CWy is defined by Eq. (1).

$$MSE(CWx, CWy) = \frac{1}{a^2} \sum_{i=0}^{a-1} \sum_{j=0}^{a-1} (CWx(i,j) - CWy(i,j))^2$$

(1)

For example, if a codebook containing 256 codewords, after performing the codebook partition, totally there are 128 divisions, every division with two codewords.

## 4. Watermark Embedding

After the codebook partition, we can start the watermark embedding process, which contains the following steps:

**Step 1:** Segment the base image (BI) into non-overlapping blocks. For each block, search the CB generated by SOFM algorithm for the closest codeword and record the index value of this codeword.

**Step 2:** Use K as the seed of the pseudorandom number generator (PRNG) and randomly pick out M*M (size of watermark image (WI)) indices generated in Step 1. Map each watermark pixel to a chosen index in sequence.

**Step 3:** In the classified codebook divisions, for every index value selected in the previous step, search for the corresponding division and codeword number and record them. When doing the recording, use S bits to keep track of the division number and use one bit to write down the codeword number, where S is set to be $\lceil \log n \rceil$. For example, if a codebook have 256 codewords where every division has two codewords, we need 7 bits (i.e., $\lceil \log (256/2) \rceil = \lceil \log 128 \rceil = 7$) to locate the division and just one bit to indicate which one of the two is the codeword. That is, in this example, S is 7. To be more specific, the index for the codeword numbered 1 in the first division is 00000001, where the first part ''0000000'' locates the division and the last bit ''1'' indicates that the codeword is numbered 1.

**Step 4:** For each recorded item, replace its corresponding number bit with the pixel value of the watermark. For example, suppose the watermark bit 1 is embedded in the recorded item 00000010, the embedding result will be 00000011 with the last bit to be replaced by the watermark bit.

**Step 5:** Find the index values corresponding to the new recorded items, and then compress the image with all of the indices. Let the compressed watermarked image be BI'.

The size of the digital watermark that can be embedded in base image depends on the size of base image and the size of the divided block.

## 5. Watermark Retrieval

To perform the watermark retrieval procedure, both the sender and the receiver possess the same codebook for the transmitted watermarked image. The receiver can recover the compressed image with the indices and retrieve the embedded watermark using the codebook. Watermark retrieval is done through the following steps:

**Step 1:** If the watermark is extracted from the VQ decoded image go to Step 2, otherwise go to Step 3.

**Step 2:** Segment the decompressed watermarked image into blocks of the same size with a*a pixels. With the help of the secret key K to perform PRNG(K), select the blocks where the watermark pixels were embedded and pick out the indices corresponding to the codewords in the codebook that are most similar to the selected blocks. Go to Step 4.

**Step 3:** With the help of the secret key K to perform PRNG(K) and select the indices embedded with the watermark pixels.

**Step 4:** In the codebook divisions, search for the divisions and codeword numbers corresponding to the selected indices. Then, recover the watermark WI' with the number bits.

## 6. Performance Analysis

In this paper, the digital watermarking technique completes the watermark embedding process in the color image through R channel during VQ encoding. However, the embedded watermark still exists in the VQ decoded image. Subjectively, human eyes can evaluate the image quality. However, the judgment is influenced easily by the factors like expertise of the viewers, experimental environments, and so on. To evaluate the image fidelity objectively, some famous measurements are adopted in this paper and described as follows.

The quality of the watermark embedded compressed image with the original base image is calculated through the peak signal-to-noise ratio (PSNR) defined as Eq. (2).

$$PSNR = 10 \log_{10} \frac{255^2}{MSE} \, dB \quad (2)$$

where MSE is the mean-square error between the original grayscale image and the distorted one. Basically, the higher the PSNR is, the less distortion there is to the host image and the distorted one.

Normalized correlation (NC) is used to judge the similarity between the original watermark WI and extracted watermark WI'. The NC is defined as Eq. (3):

$$NC = \frac{\sum_{i=0}^{M-1} \sum_{j=0}^{M-1} WI(i,j) WI'(i,j)}{\sum_{i=0}^{M-1} \sum_{j=0}^{M-1} WI(i,j)^2} \quad (3)$$

In principle, if the NC value is closer to 1, the extracted watermark is getting more similar to the embedded one.
The bit-correct rate (BCR) is computed as Eq. (4):

$$BCR (\%) = \frac{1 - \sum_{i=0}^{m} |WI - WI'|}{m} \times 100\% \quad (4)$$

The mean absolute error (MAE) is computed as Eq. (5):

$$MAE = \frac{1}{m} \sum_{i=0}^{m-1} |WI - WI'| \quad (5)$$

where *m* denote the length of the signature. Note that the quantitative index, MAE, is exploited to measure the similarity between two binary images.

## 7. Experimental Results

For showing the results of the proposed method, the 512*512*3 Lena image is used as the original image. The watermark image is a 64*64 binary bitmap. The watermark is embedded in the R component during VQ encoding. Fig 1(a), (b) and (c) shows original image, original watermark image and watermarked image respectively. From fig 1(a) and (c), it is clear that the embedding algorithm does not distort host image much and the embedded watermark still exists in the VQ decoded image with PSNR value 32.9731dB. Without attacks, the extracted watermark with the NC value as 1, MAE as 0 and BCR as 100% is shown in fig 1(d), shows the extracted watermark is still meaningful and recognizable.

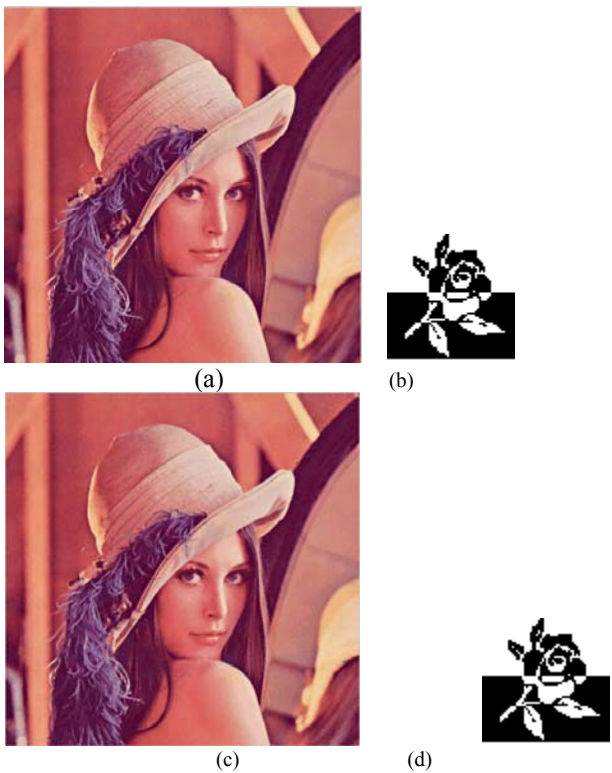

**Fig 1.** (a) Original image; (b) Original watermark;
(c) Watermarked image; (d) Extracted watermark.

In the proposed method, a codebook containing 256 codewords are obtained using SOFM algorithm and build up codebook divisions with every division containing two codewords. To bring out the effectiveness of the SOFM based VQ, it is compared with the oldest, as well as most commonly used LBG algorithm for various test images.

**Table 1.** Significance of SOFM over LBG

| Image | SOFM | | | LBG | | |
|---|---|---|---|---|---|---|
| | PSNR (dB) | CR(bpp) | Time (min) | PSNR (dB) | CR(bpp) | Time (min) |
| Rice 256X256 | 33.73 | 4.462 | 14.88 | 31.92 | 5.544 | 22.62 |
| Owl 256X256 | 26.59 | 4.482 | 14.45 | 25.95 | 5.806 | 22.41 |
| Mattface 256X256 | 32.22 | 4.448 | 13.89 | 31.61 | 5.731 | 24.72 |
| Bird 256X256 | 34.22 | 4.451 | 14.01 | 32.96 | 5.556 | 22.63 |
| Peppers 256X256 | 32.50 | 4.462 | 14.06 | 31.78 | 5.695 | 26.04 |

It is inferred from the table, that the SOFM algorithm performs better than the well-known LBG algorithm in terms of compression ratio and image quality. The effectiveness of the SOFM algorithm over the LBG algorithm for various images is plotted in Fig 2.

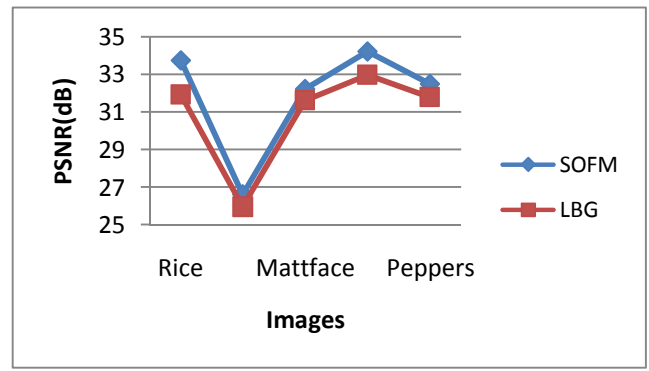

**Fig 2(a).** Performance of SOFM over LBG in terms of PSNR

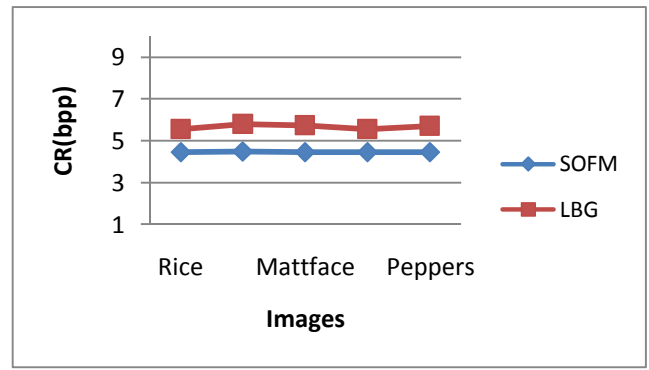

**Fig 2(b).** Performance of SOFM over LBG in terms of compression ratio

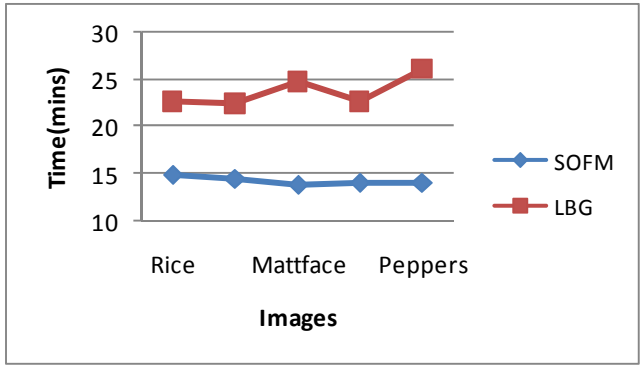

**Fig 2(c).** Performance of SOFM over LBG in terms of computation time

Comparison to the results in Fig 2(a), when we use SOFM algorithm for compression then there is a increase in PSNR value around 1-2dB (Eg., for Pepper image the PSNR value for SOFM is 32.50dB and the PSNR value for LBG is 31.78dB). In terms of compression ratio, there is an increase when we use SOFM based algorithm to compress an image. Also there is a reduction in time for computation. In fig 2(c) the computation time to generate the codebook for Pepper image using LBG algorithm is 26.04 minutes which is further reduced to 14.06 minutes in SOFM algorithm. Hence the proposed work uses SOFM algorithm for codebook generation.

### 7.1 Codebook Vs Image Quality

Fig 3 shows the effectiveness of the watermarking technique with various codebook sizes. We observe that the use of a larger codebook leads to have better watermarked image quality.

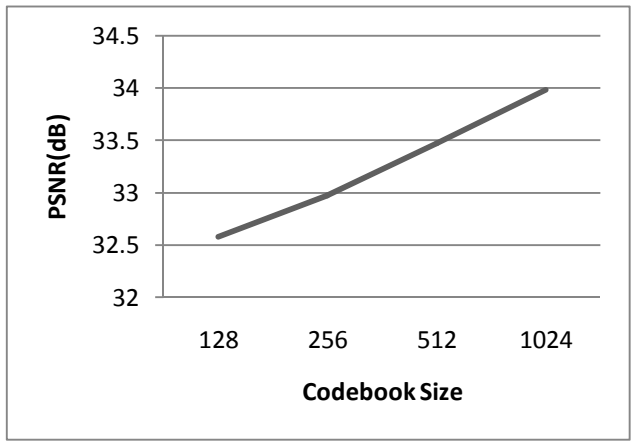

**Fig 3.** Codebook size Vs watermarked image quality

### 7.2 Filter Attacks

To measure the robustness, the watermarked image is tested by common image processing operations, geometric distortions and lossy compression. Table 2 shows the performance of the algorithm with various filter operations.

**Table 2.** Comparison of results of NC, MAE and BCR for various filter operations

| Attack Method | NC | MAE | BCR (%) |
|---|---|---|---|
| Wiener Filter (3x3) | 0.9474 | 0.1411 | 85.89 |
| Wiener Filter (5x5) | 0.8026 | 0.3384 | 66.16 |
| Wiener Filter (7x7) | 0.7541 | 0.4197 | 58.03 |
| Median Filter (3x3) | 0.9281 | 0.2280 | 77.19 |
| Median Filter (5x5) | 0.7995 | 0.3694 | 63.06 |
| Median Filter (7x7) | 0.7352 | 0.4436 | 55.64 |

Fig 4 shows the performance of algorithm against various filter operations. From the graph we observe that when the mask size increases the NC value going down. Fig 5, Fig 6 displays the effects of the cropping attacks and we observe that the extracted watermarks are obviously distinguishable by the human eye when the cropped part is filled with white pixels.

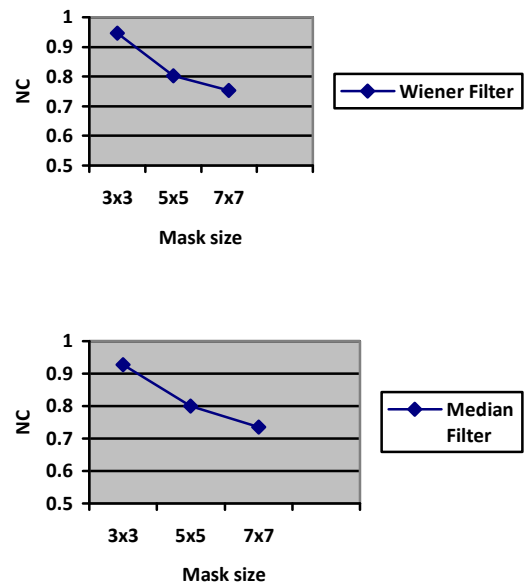

**Fig 4.** Graphs showing the performance of algorithm against various filter operations

### 7.2 Cropping and Other Attacks

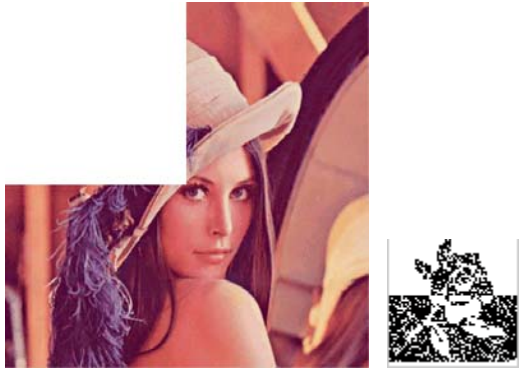

(a) Quarter cropped MAE = 0.1299

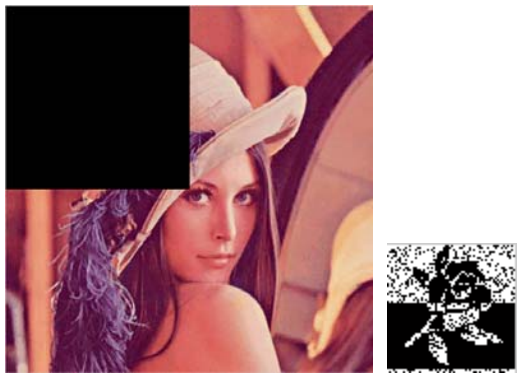

(b) Quarter cropped and filled with zero MAE = 0.1152

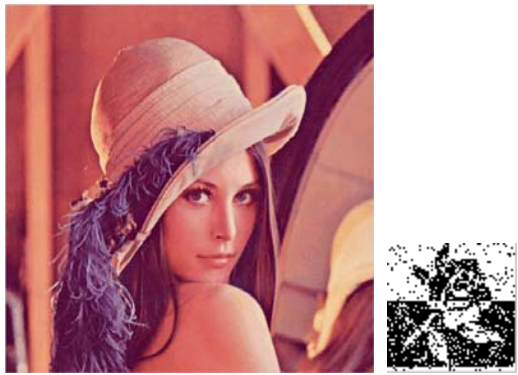

(c) Quarter cropped and replaced with original MAE = 0.1206
**Fig 5.** The quarter-cropped, watermarked image

The experimental results of other attacks also shown in table 3. It can be seen that our algorithm can successfully resist attacks by gaussian filter, sharpening, blurring, noise, enhancement, cropping in edges and interception. The watermarks are extracted with good NC, MAE and BCR values. Thus the extracted watermarks can be identified and declared correctly.

**Table 3.** Experimental results with various attacks

| Attack Method | NC | MAE | BCR (%) |
|---|---|---|---|
| Gaussian Filter | 0.9934 | 0.0127 | 98.7305 |
| Sharpening | 0.8347 | 0.1729 | 82.7148 |
| Blurring | 0.8036 | 0.4043 | 59.5703 |
| Salt & pepper noise | 0.9413 | 0.0508 | 94.9219 |
| Gaussian noise | 0.9673 | 0.0325 | 96.7529 |
| 25 rows & columns cropped in all sides | 0.8235 | 0.0935 | 90.6494 |
| Interception and deleted part filled with gray | 1 | 0.0198 | 98.0225 |
| Enhancement | 0.7245 | 0.3406 | 65.9424 |

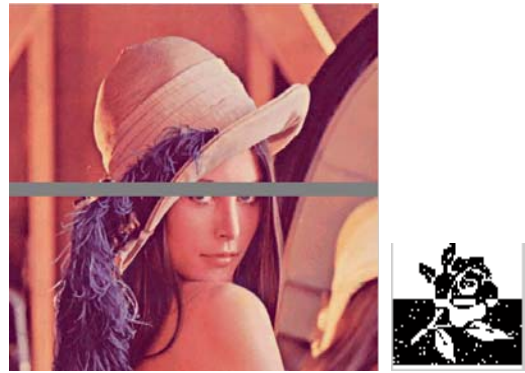

**Fig 6(a).** Simulation of intercepted and deleted index

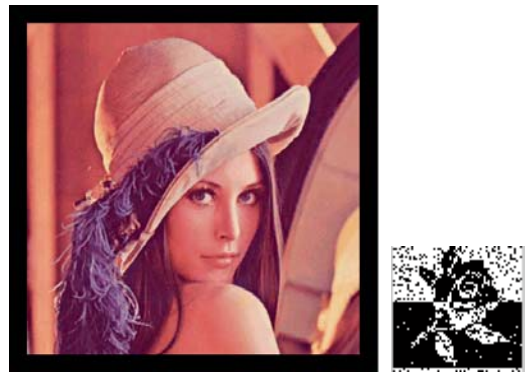

**Fig 6(b).** Simulation of cropping effect in all sides

### 7.4 JPEG Attacks

The following table 4 illustrates the influence of JPEG lossy compression on watermarked image. The results influences that our method can obtain an extracted watermark with the NC value as 1 when the quantization level is 128 and 64. Fig 7 shows the extracted watermark after the JPEG lossy compression.

**Table 4.** Experimental results of watermarks extracted from images compressed at different JPEG compression level

| Quantization level | NC | BCR (%) | PSNR (with JPEG attack) |
|---|---|---|---|
| 128 | 1 | 100 | 31.7808 |
| 64 | 1 | 96.0449 | 31.7164 |
| 32 | 0.9291 | 83.3740 | 31.3043 |
| 8 | 0.9107 | 64.4287 | 30.1549 |
| 4 | 0.8061 | 51.3184 | 28.0327 |

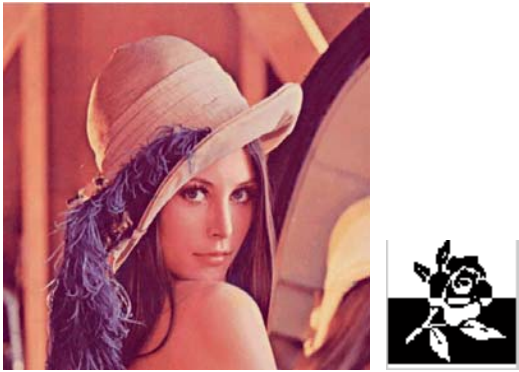

**Fig 7.** Watermark extracted from the JPEG lossy compressed image

The experiments conducted in this section proved that the images processed by our method have not only been processed by VQ but can also resist a JPEG lossy compression and other attacks. Therefore, our method can effectively protect the intellectual property of images.

## 8. Conclusion

A watermarking scheme based on SOFM in color image has been proposed in this paper for enhancing the VQ system of the watermarking ability. The simulation results illustrate that, our technique possesses the advantages of better imperceptibility, stronger robustness, faster encoding time, and easy implementation. This technique embeds the watermark in an image that has already been compressed, which saves time and space when the image is transmitted over network. Experimental results demonstrate that the proposed method is suitable for intellectual property protection applications on the Internet.

**J. Anitha** was born in Nagercoil, Tamilnadu, India, on July 14, 1983. She obtained her B.E degree in Information Technology from Mononmaniam Sundarnar University in 2004, Tirunelveli, Tamilnadu. She received her Masters degree in Computer Science and Engineering from Manonmaniam Sundarnar University in 2006, Tirunelveli, Tamilnadu. Currently she is pursuing Ph.D in the area of Image processing. Her area of interest is in the field of compression, neural network and watermarking. She is the life member of Computer Society of India.

**S. Immanuel Alex Pandian** was born in Tirunelveli, Tamilnadu, India, on July 23, 1977. He obtained his B.E degree in Electronics and Communication Engineering from Madurai Kamaraj University in 1998, Madurai, Tamilnadu. He received his Masters degree in Applied Electronics from Anna University in 2005, Chennai, Tamilnadu. Currently he is pursuing Ph.D in the area of Video processing. His area of interest is in the field of image processing, computer communication and neural network.